\documentclass[11pt,oneside,english,preprint]{amsart}
\usepackage[T1]{fontenc}
\usepackage[latin9]{inputenc}
\usepackage[a4paper]{geometry}
\usepackage{amsthm}
\usepackage{amssymb}
\usepackage{graphicx}
\usepackage{setspace}
\makeatletter

\providecommand{\tabularnewline}{\\}

\numberwithin{equation}{section}
\numberwithin{figure}{section}

\makeatother

\usepackage{babel}
\begin{document}

\title{On the Structure of Star-Polymer Networks}

\author{K. Schwenke, M. Lang and J.-U. Sommer}
\begin{abstract}
Using the bond fluctuation model we study polymer networks obtained
by endlinking of symmetric 4-arm star polymers. We consider two types
of systems. Solutions of one type (A) of star polymers and solution
of two types (A,B) of star polymer where A-type polymers can only
crosslink with B-type polymers.  We find that network defects in $A$
networks are dominated by short dangling loops close to overlap concentration
$c^{*}$. $AB$ networks develop a more perfect network structure,
since loop sizes involving an odd number of stars are impossible,
and thus, the most frequent dangling loop with largest impact on the
phantom modulus is absent. The analysis of the pair-correlation and
scattering function reveals that there is an amorphous packing of
$A$ and $B$ type stars with a homogenization of $A$ and $B$ concentrations
upon cross-linking at intermediate length scales in contrast to the
previously suggested crystalline like order of $A$ and $B$ components
at $c^{*}$ . This result is corroborated by the coincidence of the
probabilities of the shortest loop structures (which is impossible
upon the previously suggested packing of stars) in both types of networks.
 Furthermore, we derive the vector order parameters associated with
the most frequent network structures based on the phantom model. In
particular for $AB$ networks we can show that there is a dominating
cyclic defect with a clearly separated order parameter that could
be used to analyze cyclic network defects. 
\end{abstract}
\maketitle

\section{Introduction}

Polymer networks are not fully understood because of the frozen-in
disorder in the connectivity of chains, which is the result of a random
crosslinking process. The construction of networks obtained from well
defined precursor molecules and crosslinking processes is a possible
route to gain deeper understanding of the resulting polymer structures.
The research on such model networks is driven by applications where
well defined network structures composed of functional macromolecules
are needed.

Recently, T. Sakai \textit{et al.} synthesized a novel class of hydrogels
made of 4-arm star polymers with tetra-Nhydroxysuccinimide-glutarate-terminated
PEG (A-type) and tetraamine-terminated PEG (B-type) as precursor molecules,
{}``Tetra-PEG-gels'' \cite{Sakai08}. Due to the different functional
end-groups, crosslinking occurs exclusively between A- and B-type
molecules. The obtained networks show a remarkably high mechanical
strength and no excess light scattering for networks cross-linked
close to overlap concentration $c^{*}$. It was suggested \cite{Sakai08,matsunaga2009sans,matsunaga2009structure}
that these samples exhibit an extremely homogeneous network structure.

In the present work, we applied Monte Carlo simulations using the
bond fluctuation model (BFM) to study in detail two model systems:
homopolymer star networks (all reactive groups are same type $A$)
and copolymer star networks (stoichiometric mixture of $A$ and $B$
type stars that form $AB$ bonds exclusively) of four arm star-polymers
of equal arm length $N_{a}$. The goal of our work is to clarify the
reasons for the improved network structure in the above mentioned
experimental studies. In order to eliminate structural changes as
function of conversion, the crosslinking process was stopped at approximately
the same extent of reaction (95\%) for all simulations. Therefore,
significant differences in network structure of $A$ and $AB$ networks
can only arise from changes among the active material or from a different
spatial packing of the star polymers.

\begin{figure}
\includegraphics[width=0.8\columnwidth]{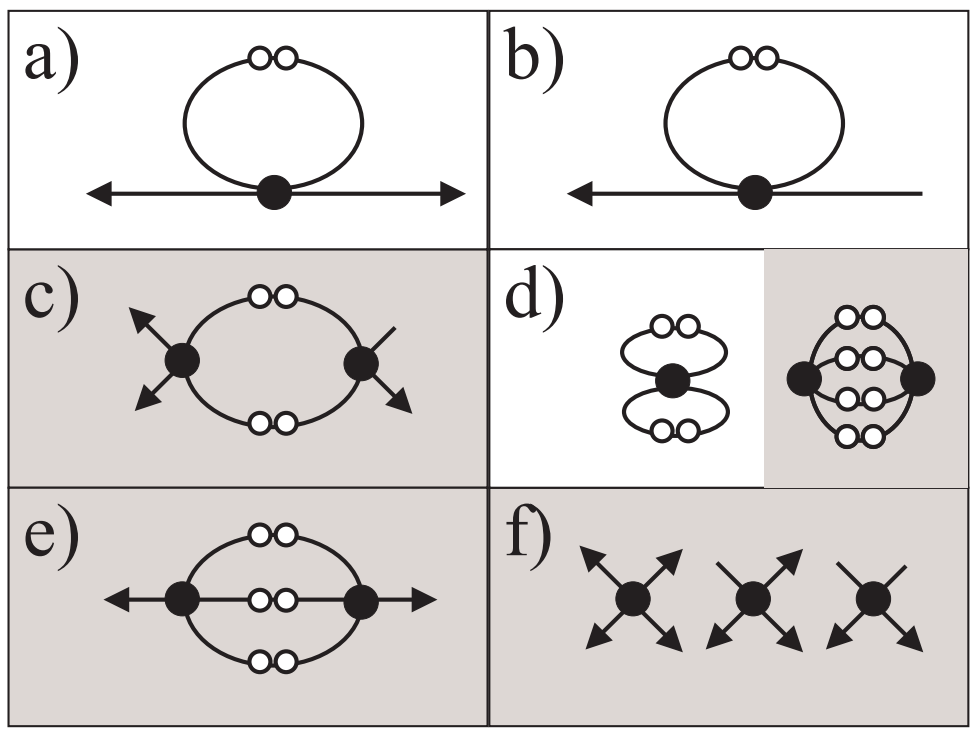}

\caption{\label{fig:Loop-structures-that-1}Loop defects in a polymer network
obtained from 4-arm star polymers. Arrows in the Figures indicate
connections to the network, filled circles symbolize star centers,
and open circles depict reacted groups. Lines, which are not terminated
by a circle or arrow display arms of the stars that do not connect
to the network. Structures with white background exist only in $A$
networks, structures with grey background are possible in $A$ and
$AB$ networks.}
\end{figure}

To consider both possibilities, we analyze the spatial order of star
polymers by pair-correlation and scattering functions before and after
cross-linking. Additionally, we analyze the predominant cyclic structures
that reduce network modulus. Figure \ref{fig:Loop-structures-that-1},
for instance, shows loop structures that are either completely inactive
as the {}``self-loop'' at a) or the full star at b). Other structures
lead to a reduced contribution \cite{flory1976statistical} of the
attached network chains to phantom modulus as the double link at c)
or the triple link at e), or can lead to an increased amount of sol
as shown at d). In the present work, we focus on exploring the structural
properties of tetra-peg star networks of Ref. \cite{Sakai08} and
discuss a possibility to detect structural defects using the concept
of segmental order parameters.

\section{Simulation Methods and Systems\label{sec:Simulation-Method-and}}

We use the bond-fluctuation model (BFM) \cite{CarmesinKremer88,DeutschBinder91}
to simulate star polymer solutions, network formation, and to determine
the properties of the obtained networks after crosslinking is terminated.
This method was chosen, since it is is known to reproduce conformational
properties and dynamics of dense polymer systems \cite{binder1995monte,wittmer2007intramolecular},
semi-dilute solutions \cite{paul1991crossover} and polymer networks
\cite{Sommer:Bfm1,LaySommerBlumen99,sommer2005segmental,lang2007trapped,LangSommer08}.
In this method, each monomer is represented by a cube occupying eight
lattice sites on a cubic lattice. The bonds between monomers are restricted
to a set of 108 bond vectors which ensure cut-avoidance of polymer
strands by checking for excluded volume. Monomer motion is modeled
by random jumps to one of the six nearest lattice positions. A move
is accepted, if the bonds connecting to the new position are still
among the set of 108 bond vectors and if no monomers overlap. All
samples were created in simulation boxes of size $L=(32MN_{a}/\phi)^{1/3}$
with periodic bondary conditions. Athermal solvent is treated implicitly
by empty lattice sites.

\begin{table}[htbp]
\begin{centering}
\begin{tabular}{|c|c|c|c|c|c|}
\hline 
$N_{a}$  & $\phi$  & $c/c^{*}$ & $M$  & $M_{sol}$ & $w_{act}$\tabularnewline
\hline 
4  & 0.116  & 0.268 & 1900  & 2 & 0.931\tabularnewline
\hline 
4 & 0.125 & 0.289 & 256 & - & -\tabularnewline
\hline 
4  & 0.140  & 0.325 & 2300  & 9 & 0.940\tabularnewline
\hline 
4 & 0.250 & 0.579 & 512 & - & -\tabularnewline
\hline 
4  & 0.375  & 0.868 & 6144  & 18 & 0.942\tabularnewline
\hline 
4  & 0.500  & 1.157 & 8192  & 16 & 0.944\tabularnewline
\hline 
8  & 0.023  & 0.104 & 1536  & 23 & 0.896\tabularnewline
\hline 
8  & 0.047  & 0.207 & 3072  & 13 & 0.930\tabularnewline
\hline 
8  & 0.063  & 0.276 & 4096  & 16 & 0.930\tabularnewline
\hline 
8  & 0.109  & 0.484 & 7168  & 12 & 0.942\tabularnewline
\hline 
8  & 0.188  & 0.829 & 1536  & 5 & 0.942\tabularnewline
\hline 
8  & 0.375  & 1.658 & 3072  & 8 & 0.945\tabularnewline
\hline 
8  & 0.500  & 2.211 & 4096  & 6 & 0.944\tabularnewline
\hline 
16 & 0.016 & 0.124 & 8 & - & -\tabularnewline
\hline 
16 & 0.023 & 0.186 & 96 & - & -\tabularnewline
\hline 
16 & 0.031 & 0.247 & 1024 & 1 & 0.938\tabularnewline
\hline 
16  & 0.047  & 0.371 & 1536  & 1 & 0.937\tabularnewline
\hline 
16  & 0.063  & 0.495 & 2048  & 7 & 0.940\tabularnewline
\hline 
16  & 0.125  & 0.990 & 4096  & 6  & 0.940\tabularnewline
\hline 
16 & 0.188 & 1.485 & 768 & - & -\tabularnewline
\hline 
16  & 0.250  & 1.979 & 8192  & 12 & 0.943\tabularnewline
\hline 
16 & 0.375 & 2.969 & 1536 & 0 & 0.942\tabularnewline
\hline 
16  & 0.500  & 3.958 & 2048  & 2 & 0.949\tabularnewline
\hline 
32 & 0.008 & 0.108 & 16 & - & -\tabularnewline
\hline 
32 & 0.016 & 0.215 & 32 & - & -\tabularnewline
\hline 
32  & 0.023  & 0.323 & 3072  & 13 & 0.932\tabularnewline
\hline 
32  & 0.031  & 0.431 & 4096  & 7 & 0.940\tabularnewline
\hline 
32 & 0.039 & 0.539 & 5120 & 6 & 0.939\tabularnewline
\hline 
32 & 0.047 & 0.646 & 96 & - & -\tabularnewline
\hline 
32 & 0.063 & 0.862 & 128 & 0 & 0.944\tabularnewline
\hline 
32  & 0.094  & 1.292 & 1536  & 1 & 0.944\tabularnewline
\hline 
32  & 0.125  & 1.723 & 2048  & 5 & 0.944\tabularnewline
\hline 
32  & 0.188  & 2.585 & 3072  & 0 & 0.947\tabularnewline
\hline 
32  & 0.250  & 3.446 & 4096  & 9 & 0.944\tabularnewline
\hline 
32 & 0.375 & 5.169 & 768 & - & -\tabularnewline
\hline 
32  & 0.500  & 6.892 & 8192  & 7  & 0.945\tabularnewline
\hline 
64 & 0.016 & 0.370 & 128 & - & -\tabularnewline
\hline 
64 & 0.023 & 0.555 & 192 & - & -\tabularnewline
\hline 
64 & 0.031 & 0.741 & 256 & - & -\tabularnewline
\hline 
64 & 0.047 & 1.111 & 384 & - & -\tabularnewline
\hline 
64 & 0.125 & 2.962 & 1024 & 1 & 0.946\tabularnewline
\hline 
64  & 0.188  & 4.443 & 1536  & 1 & 0.943\tabularnewline
\hline 
64  & 0.250  & 5.924 & 2048  & 0 & 0.947\tabularnewline
\hline 
64  & 0.375  & 8.886 & 3072  & 1  & 0.946\tabularnewline
\hline 
64  & 0.500  & 11.85 & 4096  & 0 & 0.947\tabularnewline
\hline 
\end{tabular}
\par\end{centering}

\caption{Copolymer {}``$AB$'' solutions and networks: The polymer volume
fraction $\phi$, the arm length $N_{a}$, and the number $M$ of
stars in the system describe the star polymer solutions of Fig. \ref{fig:scaling}.
The solutions were crosslinked up to the extent of reaction $p\approx0.95$
with the number of stars in sol $M_{sol}$, the weight fraction of
active material $w_{act}$, if this information is provided in the
table.}

\label{tab:simsystems} 
\end{table}

\begin{table}[htbp]
\begin{centering}
\begin{tabular}{|c|c|c|c|c|c|}
\hline 
$N_{a}$  & $\phi$  & $c/c^{*}$  & $M$  & $M_{sol}$ & $w_{act}$\tabularnewline
\hline 
32  & 0.016  & 0.215 & 2048  & 178 & 0.594\tabularnewline
\hline 
32  & 0.023  & 0.323 & 3072  & 70 & 0.705\tabularnewline
\hline 
32  & 0.031  & 0.431 & 4096  & 43 & 0.755\tabularnewline
\hline 
32  & 0.063  & 0.862 & 1024  & 2 & 0.858\tabularnewline
\hline 
32 & 0.094 & 1.292 & 1536 & 4 & 0.878\tabularnewline
\hline 
32  & 0.125  & 1.723 & 2048  & 3 & 0.888\tabularnewline
\hline 
32 & 0.188 & 2.585 & 3072 & 0 & 0.901\tabularnewline
\hline 
32  & 0.250  & 3.446 & 4096  & 1 & 0.908\tabularnewline
\hline 
32  & 0.500  & 5.169 & 8192  & 5 & 0.921\tabularnewline
\hline 
\end{tabular}
\par\end{centering}

\caption{Homopolymer networks $A$. Notation as in table \ref{tab:simsystems}.}

\label{tab:simsystemsAA} 
\end{table}

Monodisperse solutions of star polymers with 4 arms were created as
described in table \ref{tab:simsystems} and \ref{tab:simsystemsAA}.
The monomer volume fractions span the range from dilute solutions
up to concentrated systems at $\phi=0.5$ comparable to polymer melts
\cite{paul1991crossover}. Each star contains a ring of 4 monomers
as core with $N_{a}-1$ arm monomers attached. The polymer solutions
were relaxed over a period of several relaxation times of the stars
as checked by mean square displacements of full star polymers and
end-to-end vector auto-correlation of star arms. Reaction took place
(i.e. a permanent bond is introduced) whenever two previously unreacted
chain ends of type $A$ (for $A$-type networks) or one of type $A$
and one of type $B$ (for $AB$-type networks ) hit each other during
the course of their motion at minimum separation on the lattice. This
criterion was used in order to avoid the formation of bonds that can
no longer move%
\footnote{For instance, the formation of bond (3,1,0) from the origin, if a
neighboring bond (1,-3,0) starts from position (1,2,0)%
}. Reaction was stopped at about 95\% of maximum possible extent of
reaction in order to eliminate structural changes as function of conversion.
The properties of the solutions before cross-linking and of the networks
after cross-linking are evaluated in the following sections.

\section{Solutions of Star Polymers\label{sec:Solutions-of-Star}}

In the following, we denote the number concentration of monomers as
$c$. We define the overlap concentration of a monodisperse polymer
solution geometrically as 
\begin{equation}
c^{*}=\frac{3N}{4\pi R_{g0}^{3}}\sim b^{-3}N^{1-3\nu}\,\,.\label{eq:1}
\end{equation}
Here, $b$ denotes the root mean square average bond length, $N$
the degree of polymerization, $R_{g0}\sim bN^{\nu}$ the radius of
gyration of dilute polymers, and the exponent $\nu\approx0.588$ for
long chains in athermal solvent. For ideal $f$-arm stars we have
$\nu=0.5$ and thus, $R_{g}^{2}=(3-2/f)\cdot N_{a}b^{2}/6$. We focus
on $c/c^{*}$ being the scaling variable for semi-dilute solutions,
since we expect the same number of neighboring molecules for a given
polymer at same $c/c^{*}$, and thus, identical amounts of cyclic
network defects or, if existing, a similar spatial ordering or packing
of the stars. The amount of linear or branched dangling material is
similar among all networks, since the reactions were stopped at the
same extent of reaction (95\%) and the $AB$ networks are stoichiometric
mixtures of stars with identical architecture but different reactive
groups.

Below $c^{*}$, the chain conformations are comparable to an isolated
coil in athermal solvent \cite{Degennes}. Semi-dilute solutions with
$c>c^{*}$ can be considered as divided into space-filling correlation
volumes $\xi^{3}$, called blobs, of polymer concentration $c\approx gb^{3}/\xi^{3}$
containing $g$ monomers each. The blob size decreases with concentration
as $\xi\sim bc^{-\nu/(3\nu-1)}$. Since the total coil conformation
can be considered as random walk of blobs of size $\xi$, the square
chain size for $c>c^{*}$ is expected to decrease as 
\begin{equation}
R_{e}^{2}\approx\xi^{2}N/g\sim b^{2}(c/c^{*})^{(1-2\nu)/(3\nu-1)}N^{2\nu}\label{eq:2}
\end{equation}
with $(1-2\nu)/(3\nu-1)\approx-0.23$. Note that $R_{g}^{2}\sim R_{e}^{2}$
and that the above result was originally derived for linear chains
\cite{Degennes}. It can be expected to hold for 4-arm stars of our
study, since additional regimes proposed by Daoud \cite{daoud1982star}
require a larger number of arms. 

\begin{figure}
\begin{center} \includegraphics[angle=270,width=1\columnwidth]{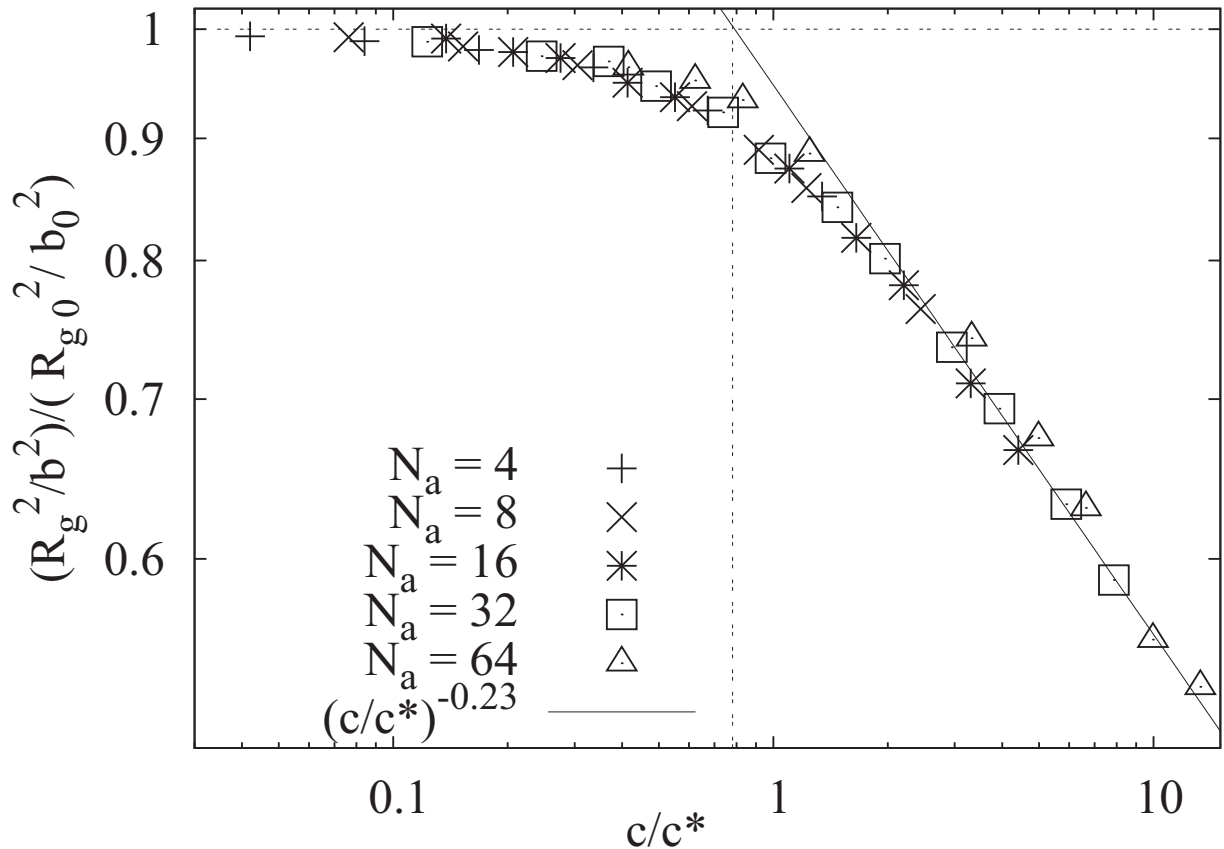}\end{center}
\caption{\label{fig:scaling} Scaling plot of the square radius of gyration
$R_{g}^{2}/b^{2}$ of star polymers normalized by the square size
of isolated stars $R_{g0}^{2}/b_{0}^{2}$ as function of $c/c^{*}$
for a series of star polymer solutions of different arm length $N_{a}$
and concentrations $c$. The asymptotic behaviors are indicated by
straight lines, their intersection point can be used to define the
value of $c*$.}
\end{figure}

Equation (\ref{eq:2}) is compared with the simulation data at Figure
\ref{fig:scaling}. For analysis, the concentration dependence of
the mean square bond length $b^{2}$, cf. Ref. \cite{paul1991crossover},
is corrected by computing $R_{g}^{2}(c)/b^{2}(c)$ for each particular
concentration $c$. The data collapses roughly on a single curve with
a cross-over region of about one decade when plotting normalized chain
size as function of $c/c^{*}$. Note that using a sphere of $R_{g0}$
for defining $c^{*}$ is nearly quantitative when comparing the intersection
point of the scaling of $R_{g}^{2}(c/c*)$ for $c\gg c*$ and for
$c\ll c*$ in Figure \ref{fig:scaling}. Therefore, we use $c^{*}$
as defined in equation \ref{eq:1} as reference for the analysis below.
We conclude that 4-arm flexible star polymers obey concentration scaling
as derived for linear polymers.

\section{Spatial order of star polymers\label{sec:Network-Structure}}

\begin{figure}[htb]
\begin{center}\includegraphics[angle=270,width=1\columnwidth]{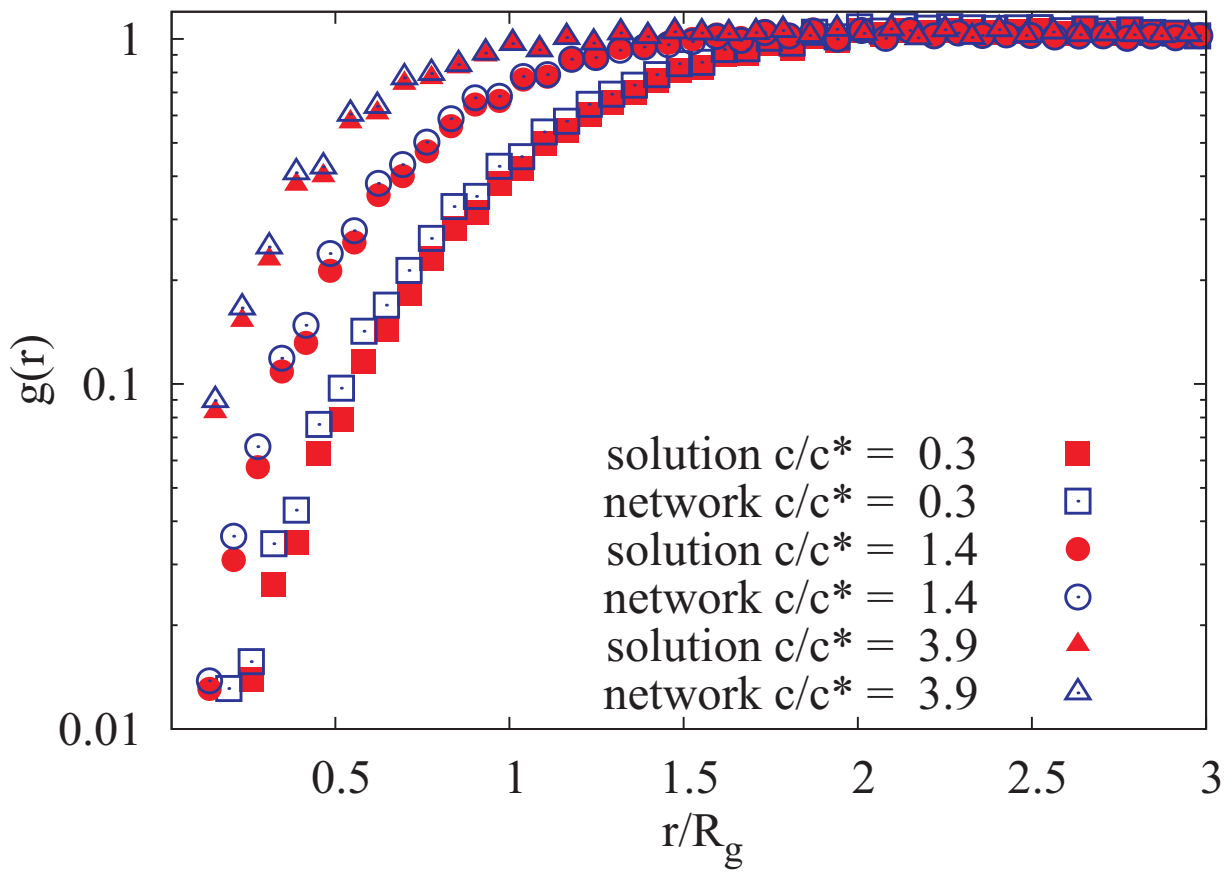}\end{center}

\caption{\label{fig:corr-1} Pair correlation function of central monomers
of different stars for networks (full symbols) and solutions (open
symbols) with $N_{a}=32$ as function of the normalized distance $r/R_{g}$
for different polymer concentrations close to $c^{*}$.}
\end{figure}

Star polymers, in particular with short arms, might repel each other
to a larger extent as linear chains. This could lead to some spatial
order similar to a hard-sphere like packing as proposed in Ref.~\cite{Sakai08}.
In Fig.\ref{fig:corr-1} we display the the pair correlation function
$g(r)$ of all stars centers. Our results show the formation of a
depletion zone of center monomers of other stars around the center
of a given star, which remains almost unmodified upon cross-linking.
The shape of this depletion zone does not correspond with results
of hard-core fluids that typically show a sharp depletion with an
oszillating correlation function at distances larger than particle
diameter \cite{yarnell1973}. In contrast, it is rather consistent
with the correlation hole as typically observed in polymer solutions
or dense melts \cite{wittmer2007intramolecular}. Cross-linking leads
only to a weak decrease of the depletion width due to attractive forces
along the bonds among connected stars. The formation of additional
peaks indicating long range spatial order cannot be observed.

\begin{figure}
\includegraphics[angle=270,width=1\columnwidth]{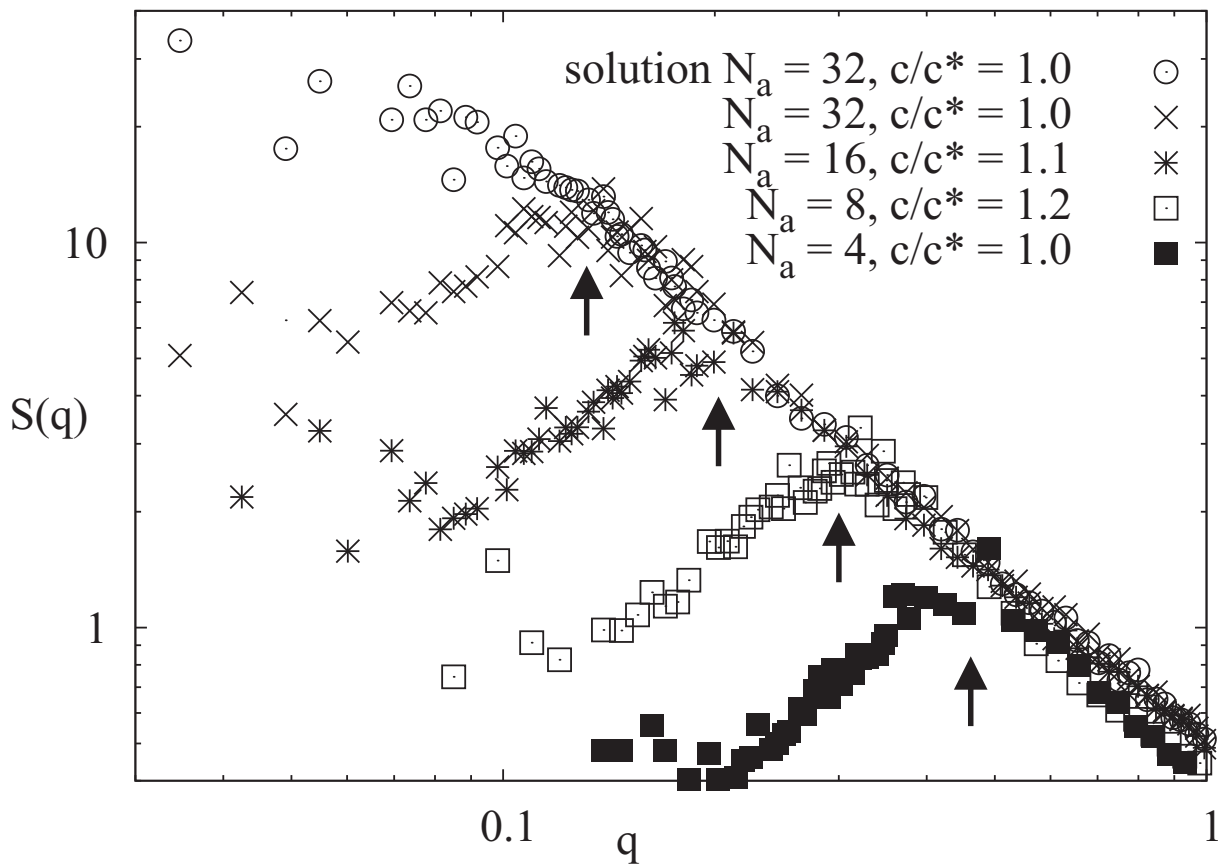}

\caption{\label{fig:Selective-structure-factor-1-1}Selective structure factor
of $A$ type stars in $AB$ networks and solutions. The arrows indicate
the peak position $q_{peak}$ of the over-next stars along the structure,
which are the next connected neighbour of same type.}
\end{figure}

The collective structure factor $S(\mathbf{q})$ as obtained in typical
scattering experiments is the Fourier-transform of the pair correlation
function 
\begin{equation}
S(\mathbf{q})=\int\exp(-i\mathbf{qr}\left(g\left(\mathbf{r}\right)-\left\langle c_{m}\right\rangle \right)\mbox{d}^{3}\mathbf{r}\,.\label{eq:S(q)}
\end{equation}
From the analysis of the pair correlation function at Fig.~\ref{fig:corr-1}
it is already obvious that the \emph{non-selevtive }(with respect
to $A$ and $B$ stars) structure factors including all star polymers
are almost indistinguishable before and after cross-linking (data
not shown). The situation is different for the \emph{selective structure
}factors as computed for all monomers of only one species of stars
inside the $AB$ networks, see Figure \ref{fig:Selective-structure-factor-1-1}.
The data shows a peak at the average distance of the over-next star
that is located about four arm lengths apart from a given star center.
No higher order peaks can be resolved, which indicates the absence
of long-range order. We note that the selective scattering data of
$AB$ solutions coincides with the corresponding network data of Figure
\ref{fig:Selective-structure-factor-1-1} for $q>q_{peak}$ as shown
for $N_{a}=32$. For $q<q_{peak},$ the solution data show an Ornstein-Zernike
type crossover and display higher scattering as compared to network
data. We explain this observation by the fact that network formation
at large conversion $p$ induces reduced density fluctuations of $A$
vs. $B$ type stars at intermediate lengths. Note that this observation
is not in contrast to the usually observed excess scattering in polymer
networks at equilibrium swelling or as function of the swelling ratio
\cite{bastide1996physical,LangSommer08}, since here, we analyze the
scattering of networks at preparation conditions. Note that the selective
structure factor for $q<q_{peak}$ is clearly above the overall structure
factor of both types of stars (not included in Figure \ref{fig:Selective-structure-factor-1-1}).
Based on our simulation results we conclude that this cross-linking
induced order is much weaker than the repulsion among stars inside
the sample and thus, does not lead to the formation of lattice-like
structures.

In summary, we can only confirm soft repulsion among stars but do
not find any indication of a spatial ordering beyond a weak reduction
of density fluctuations of different type of stars inside the network
as compared to $AB$ solutions. Therefore, we now focus on network
connectivity.

\section{Loop defects inside the network structure\label{sec:Small-loops-inside}}

An ideal model network of any functionality can be imagined by considering
an infinitely branching structure like a Bethe lattice of same functionality,
as used to derive the phantom modulus \cite{Rubinstein}. An ideal
network structure of functionality four, for instance, can also be
visualized by a diamond lattice like connectivity \cite{Sakai08}.
As compared to such ideal connectivities, the random crosslinking
process always leads to the formation of various defects as short
loops, dangling network strands, or more complex inactive structures.
Any of these defects diminishes the elastic modulus of the networks.
However, for networks of functionality $f=4$ at high conversion it
is known that the fraction of complex inactive structures is decreasing
exponentially with size \cite{lang2005network}. The amount of linear
dangling material is nearly constant for star polymers, if all samples
were crosslinked up to the same conversion close to completion. Therefore,
the remaining main difference in the connectivity of star networks
close to $c^{*}$ at high conversion must be with respect to the formation
of short cyclic structures.

The simplest loop-like defects and stars with dangling strands are
sketched in Fig.~\ref{fig:Loop-structures-that-1}. Below we use
the following notation to distinguish different star connectivities
(cf. Figure \ref{fig:Loop-structures-that-1}): $R_{i}^{j}$ is used
to denote stars that are part of $j$ ring structures containing each
$i$ molecules. If $j$ is missing, it is equal to 1. $I_{j}$ is
used to denote $j$ single ({}``ideal'') connections to different
stars. Thus, the structures of Fig. \ref{fig:Loop-structures-that-1}
are denoted as a) $R_{1}I_{2}$, b) $R_{1}I_{1}$, c) both stars are
$R_{2}^{2}$, d) left star is $R_{1}^{2}$ and both stars on the right
$R_{2}^{3}$, e) the left star is $R_{2}I_{2}$, while the star on
the right $R_{2}I_{1}$, f) from left to right $I_{4}$, $I_{3}$,
$I_{2}$. Note that conformations are written in italics, while the
total fraction of monomers in loops of $i$ molecules is denoted as
$\mbox{R}_{i}$.

For a simple approximative treatment of short loop structures in networks
we use the results of Ref. \cite{lang2005intramolecular} and refer
the reader to this work for more details. There, the following approximations
are made: equal reactivity among the functional groups, homogeneous
samples, no effects of excluded volume on the spatial arrangment of
reactive groups and no effects of smaller loops onto the formation
of larger ones. Then, the rate of ring formation of short loops in
irreversible linear polymerization reactions can be approximated as
: 
\begin{equation}
\frac{\mbox{d}\mbox{R}_{i}}{\mbox{d}p}\approx p^{i-1}(1-p)\cdot\frac{c_{int,i}}{c_{int,i}+c_{j\neq i}}\ .\label{eq:Loop_general}
\end{equation}
Here, $p$ is the extent of reaction and $\mbox{R}_{i}$ the amount
of rings containing $i$ molecules. For computation, one reaction
partner is considered to be at the origin. The concentration $c_{int,i}$
is the concentration of the second reaction partner at the origin
whereby this molecule is a minimum of $i$ molecules along the connective
structure apart. $c_{j\neq i}$ is the concentration at the origin
of all other reactive groups not being $i$ connections apart. For
overlapping molecules one typically has $c_{int,i}\ll c_{j\neq i}\approx c_{ext,0}\cdot(1-p)$
with $c_{ext,0}$ the initial concentration of not-attached {}``external''
reactive groups. 

For network forming reactions, $p^{i}$ is replaced by a branching
term $[p(f-1)]^{i}$ for $A$ networks and $[p_{A}(f-1)p_{B}(g-1)]^{j}$
for $AB$ networks \cite{lang2005intramolecular} that counts the
average number of reactive sites attached $i$ or $j=2i$ molecules
apart. Since for our series of simulations $p_{A}=p_{B}=p$ and $g=f$,
the $AB$ term reduces to $[p(f-1)]^{2i}$ indicating that only \emph{even}
ring sizes can be realized in $AB$ networks. 

The concentration of attached groups $c_{int,i}$ is estimated using
the blob picture for chain conformations in semi-dilute solutions:
a chain performs a random walk of concentration blobs of size $\xi$
with $g$ monomers per blob and $N/g$ blobs per chain. The return
probability for this random walk is given by 
\begin{equation}
\Phi(0,N/g)=\left(\frac{3}{2\pi\frac{N}{g}\xi^{2}}\right)^{3/2}.\label{eq:loop-gauss0}
\end{equation}
Since $\xi^{2}/g\approx c^{(1-2\nu)/(3\nu-1)}\approx c^{-0.23}$ in
the athermal case and $N=2N_{a}$ for stars we find for a single reactive
site 
\begin{equation}
\Phi(0,iN_{a})\approx\left(\frac{3}{4\pi iN_{a}c^{-0.23}}\right)^{3/2}\approx0.12\cdot(iN_{a})^{-3/2}c^{0.35}.\label{eq:concentration}
\end{equation}
For $A$ networks we have $c_{int,i}\approx[p(f-1)]^{i}\Phi(0,iN_{a})$
and we can integrate equation (\ref{eq:Loop_general}) using the approximation
$fc=c_{ext,0}\gg c_{int,i}$ to obtain 
\begin{equation}
\mbox{R}_{i}\approx\frac{0.12}{if}[p(f-1)]^{i}(iN_{a})^{-3/2}c^{-0.65}\label{eq:loop-2}
\end{equation}
as prediction for the fraction of rings of size $2iN_{a}$ in star
polymer networks. For $AB$ networks, the prefactor changes from 0.12
to $0.24$ for even $i$, since only half of all reactive groups are
possible reaction partners. 

Note that $c^{*}$ is not the limiting concentration for network formation.
Network formation is still possible at concentrations at which at
least 1 out of $f-1$ bonds connects to at least one other molecule.
Let $x$ denote the number of reactive groups attached to other stars
in the pervaded volume of a given star. Then, $x/(f-1+x)$ is roughly
the probability to connect to a different star. Since each star has
$f-1$ attempts to connect we obtain from $x(f-1)/(f-1+x)=1$ that
$x=(f-1)/(f-2)$. Since connections are possible within the range
$R_{e}$ around the star center, this leads to a limiting polymer
concentration of 
\begin{equation}
c_{net,A}^{*}\approx\frac{f-1}{f-2}\cdot\frac{3N_{a}}{4\pi R_{e}^{3}}<c^{*}\label{eq:cnetA}
\end{equation}
 for $A$ networks, which is about one order of magnitude below $c^{*}$
for $f=4$. 

For $c<c^{*}$ and ignoring the effect of fluctuations, the return
probability of equation (\ref{eq:concentration}) becomes concentration
independent and thus, 
\begin{equation}
\mbox{R}_{i}\sim c^{-1}.\label{eq:below overlap}
\end{equation}

\begin{figure}[htbp]
\includegraphics[angle=270,width=1\columnwidth]{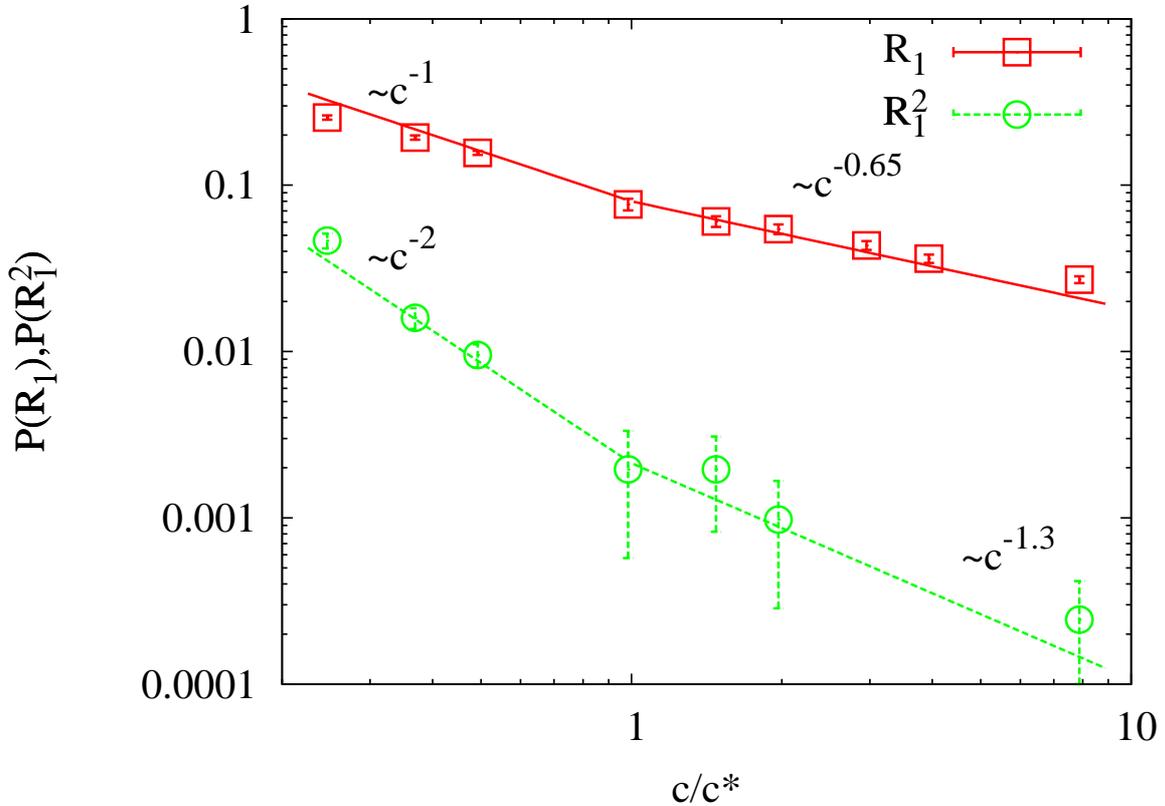} \caption{Number fraction $P(\mbox{R}_{1})$ of self loops normalized to total
number of possible bonds between stars and number fraction of stars
$P(R_{1}^{2})$ forming two self loops $R_{1}^{2}$ in the reaction
container. }

\label{fig:selfloops} 
\end{figure}

Figure \ref{fig:selfloops} shows the concentration dependence of
the number fractions of self-loops $P(\mbox{R}_{1})=\mbox{R}_{1}/(Mf/2)$
among all bonds (which is the number fraction of bonds {}``wasted''
in dangling loops) and the number fraction of stars forming two self
loops $R_{1}^{2}$. We find $P(R_{1}^{2})=\mbox{R}_{1}^{2}/M=P(\mbox{R}_{1})^{2}/3$
at $p\approx1$, since there are $f(f-1)/2$ distinguishable ways
to form the first $R_{1}$ and only one for the second, and there
are $f/2$ bonds per star (different normations). In order to show
this dependence, we fitted the more accurate $P(\mbox{R}_{1})$ data
by equation \ref{eq:loop-2} and computed from this fit the predictions
for the $P(R_{1}^{2})$. Note that this kind of procedure is only
possible, if the over all loop fraction is small as compared to the
remainder of the network structure. For larger loop fractions one
has to use a more detailed approach that also explains the deviations
for $\mbox{R}_{1}$ at low concentrations \cite{Schwenke2}. The $R_{1}^{2}$
data at the lowest concentration shows a stronger than predicted dependence
on concentration. This can be explained by concentration fluctuations
of $A$ molecules and the rapid reaction in our simulations, because
isolated stars can only form $R_{1}^{2}$ structures without collisions
with other stars until full conversion. The data at $P(R_{1}^{2})\approx10^{-3}$
corresponds to a very limited number of sol molecules with absolute
counts of $R_{1}^{2}$ on order unity (cf. Table \ref{tab:simsystemsAA}).
Therefore, the missing two data points are due to samples, for which
no $R_{1}^{2}$ could be detected. 

\begin{figure}
\includegraphics[angle=270,width=1\columnwidth]{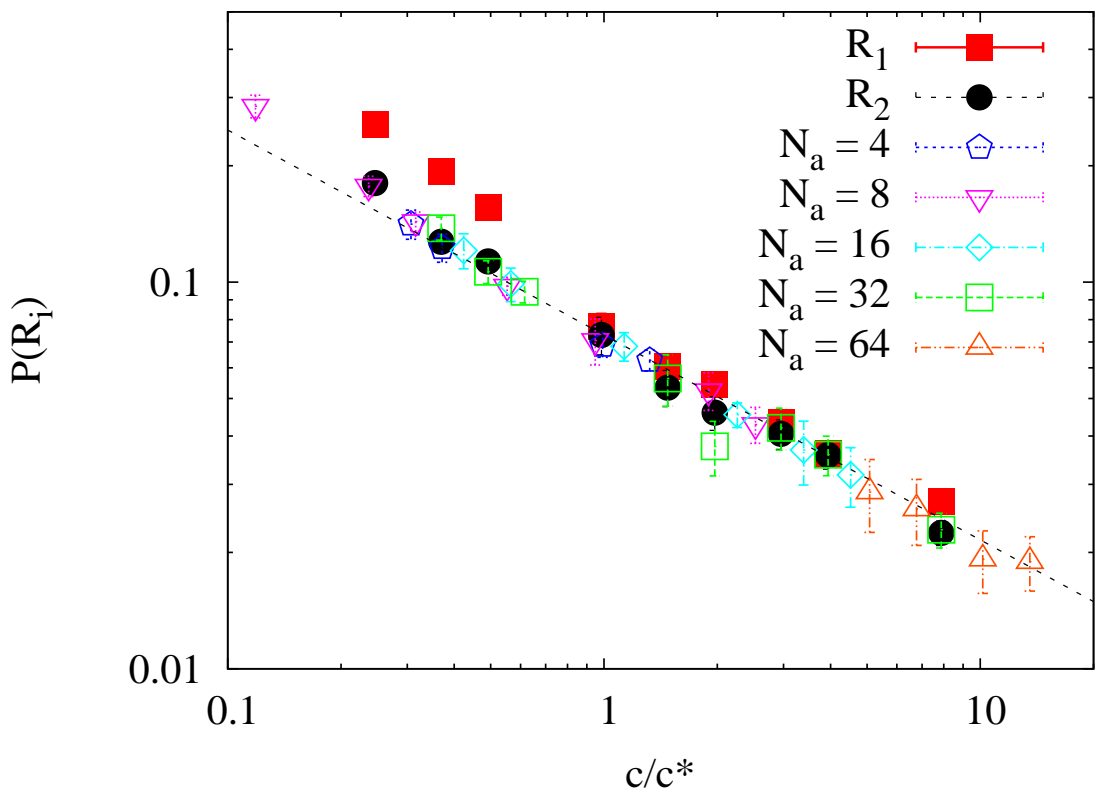}

\caption{The plot shows the number fractions of dangling links $P(\mbox{R}_{1})$,
of $A$ networks (full circles) and the number fractions of double
links $P(\mbox{R}_{2})$ data of $A$ (full squares) and $AB$ networks
(open symbols). The $P(\mbox{R}_{2})$ data of $A$ networks is multiplied
by a factor of $2/(1-P(\mbox{R}_{1}))$ as explained in the text.
The dashed line is a best fit to a power law for the $P(\mbox{R}_{2})$
data at $c\gtrsim c^{*}$.}

\label{fig:doubles} 
\end{figure}

From equation (\ref{eq:loop-2}) it can be found that $R_{i}/R_{j}\approx[p(f-1)]^{|i-j|}(j/i)^{5/2}$
independent of concentration, thus $\mbox{R}_{2}/\mbox{R}_{1}\approx1/2$
for the parameters of our simulations. The density of one type of
reactive groups in $AB$ networks is half the density as the $A$
groups in $A$ networks at same concentration, which doubles the probability
for ring formation in $AB$ networks. Thus, the dangling ring data
$P(\mbox{R}_{1})$ for $A$ networks is expected to collapse onto
the double link data $P(\mbox{R}_{2})$ of $AB$ networks. The same
holds for $P(\mbox{R}_{2})$ of $A$ networks, if the data is multiplied
by a factor of two and corrected for the amount of bonds between stars
$P(\mbox{R}_{1})$, which cannot form rings $\mbox{R}_{2}$, because
they are incorporated into rings $\mbox{R}_{1}$. The good agreement
among all data can be seen from the collapse of all $\mbox{R}_{2}$
and $\mbox{R}_{1}$ data in Figure \ref{fig:doubles} for $c>c^{*}$.
Below $c^{*}$ we note that connected pairs, triples etc ... of stars
(as neccesary for loops $\mbox{R}_{i}$) behave like increasingly
larger molecules with decreasing $c^{*}$. Therefore, a simple scaling
as function of the overlap concentration $c^{*}$ of individual stars
does no longer work.

A best fit (dashed line) of the data in Figure \ref{fig:doubles}
yields $P(\mbox{R}_{2})\approx0.073(c/c^{*})^{-0.53}$, which is a
slighly smaller power than the predicted 0.65. The absolute amount
of loops is about 30\% less than predicted from equation (\ref{eq:loop-2})
at $c^{*}$. The corrections to scaling are due to the neglect of
changes in the shape of the correlation hole of the stars and a corresponding
shift of the average positions of the end monomers by using the above
simple scaling approach. Furthermore, at $c^{*}$ the approximation
$c_{int,i}\ll c_{j\neq i}$ no longer works.

The above results can be used to understand the different stability
of $A$ and $AB$ networks by considering the different impact of
loops $R_{1}$ and $R_{2}$ onto the network structure. When comparing
$I_{4}$ with $R_{1}I_{2}$ we find that for $R_{1}I_{2}$ 50\% of
the polymer is fully removed of the active network, while the remaining
network strand is doubled in length. When comparing $I_{4}$ with
$R_{2}I_{2}$ we find that still all polymer is active, while the
effective functionality of the cross-link is only reduced by one.
This clear difference can be seen by the strong impact of concentration
onto the weight fraction of the active material at table \ref{tab:simsystemsAA},
while the data of the $AB$ network at table \ref{tab:simsystems}
is almost independent of concentration for the parameters of our study.

Altogether we find that $A$ and $AB$ networks show exactly the same
scaling for the amounts of short loops at $c>c^{*}$ after correcting
for the concentrations of reactive groups. We note that the observed
behaviour for short loops is not in agreement with assuming a diamond
lattice like network structure close to $c^{*}.$ The most important
difference between both types of networks is the absence of loops
$R_{1}$ (self-loops) in $AB$ networks. The frequent occurence of
this type of defects in $A$ networks, however, leads to a substantial
decrease in the volume fraction of active material in $A$ networks
as compared to $AB$ networks.

\section{Segmental order parameters and network defects}

Computer simulations allow to measure directly vector and tensor order
parameters in a polymer network \cite{sommer2005segmental}. In this
section we explore the relations between defects in network structure
and segmental order parameter. Since the vector order parameter requires
much less sampling time as compared to the tensor order parameter
\cite{sommer2005segmental}, we restrict our discussions to the vector
order parameter in the following.

Let $N=2N_{a}$ denote in this section the number of monomers between
two connected star centers. The vector order parameter $m_{k}$ of
segment $k=1,...,N-1$ along this chain is defined via the long time
limit $t\rightarrow\infty$ of the autocorrelation function 
\begin{equation}
m_{k}(t)=\left\langle \mathbf{n}_{k}(0)\cdot\mathbf{n}_{k}(t)\right\rangle \label{VOP-def}
\end{equation}
with $\mathbf{n}_{k}=\mathbf{b}_{k}/\left\langle \mathbf{b}_{k}^{2}\right\rangle ^{1/2}$
being the normalized segment vector and $\mathbf{b}_{k}$ denotes
the actual segment vector with monomer index $k$. For ideal chains
of $N$ segments with the ends fixed at distance $R$ we have for
each $k$ 
\begin{equation}
m_{k}=\frac{R^{2}}{b^{2}N^{2}}.\label{vop single chain}
\end{equation}
Thus, for set of ideal chains with the ends fixed according to a Gaussian
end-to-end distribution we obtain for the ensembe average of all chains
and order parameters (as indicated by square brackets $\left[...\right]$)
\begin{equation}
\left[m_{k}\right]=\frac{\left[R^{2}\right]}{b^{2}N^{2}}=\frac{b^{2}N}{b^{2}N^{2}}=\frac{1}{N}.\label{eq:VOP}
\end{equation}

Here, we note that the effect of excluded volume on the vector order
parameter is entirely determined by the change in chain extension
in contrast to the tensor order parameter~\cite{sommer_PRE08}. Most
samples of our simulation series are in the vicinity of $c^{*}$ and
are built of weakly entangled stars. Therefore, we will use the phantom
model to obtain a simplified theoretical prediction for the order
parameter of different network structures. This prediction can only
be applied close to or slightly below $c^{*}$.

Our calculations are based on the following simplifications: The phantom
model can be reduced to the affine model, by computing the corresponding
combined chain $N_{comb}$ with fixed ends that describe the deformations
of the network strands \cite{Rubinstein}. To this end we assume a
network structure similar to the Bethe lattice except of one single
defect. The vector order parameter is then computed analogous to the
derivation of phantom modulus. Details are shown in the Appendix for
connectivities $I_{4}$ and the double links $R_{2}$ of connectivity
$R_{2}I_{2}$. Similar to these examples we also computed combined
chains and order parameters of the most abundant network structures.
The results are summarized at Table \ref{tab:Combined-chains-}. In
order to highlight the importance of the particular structure, we
also included the measured fraction of polymer with the particular
combined chains in the different structures%
\footnote{For instance in the structure $R_{1}I_{2}$ there is only 50\% of
the star not dangling; but since each of the two connected stars also
contribute one arm that is part of the combined chain, there is a
fraction of 12\% of network polymer that is part of the combined chain
indicated at position n, while there is 6\% polymer in dangling rings.
The connectivity $R_{1}I_{2}$ in $A$ networks was included for illustrating
the fact that the loop $R_{1}$ itself cannot be detected, since it
is part of the dangling material. The strand attached to this loop
has some potential for analysis due to an order parameter of $3N$,
which is, however, clearly harder to distinguish of $2N$ as the order
parameter $4N$ of $R_{2}I_{2}$.%
} at concentration $c^{*}$. Note that we neglected corrections for
the amount of active material for simplicity, since at or above $c^{*}$
almost all exisiting connections in $AB$ networks at high conversion
are part of the active material (cf. table \ref{tab:simsystems}:
about 94\% of star arms is active, whereby 95\% of arms are connected). 

\begin{table}
\begin{tabular}{|c|c|c|c|c|c|}
\hline 
connectivity & pos & \% $A$ & \% $AB$ & $N_{comb}$  & $m^{-1}$ \tabularnewline
\hline 
\hline 
$I_{4}$ &  & 57 & 58 & $2N$ & $2N$\tabularnewline
\hline 
$R_{2}I_{2}$ & r & 5.5 & 11 & $2N$ & $4N$\tabularnewline
\hline 
-''- & n & 5.5 & 11 & $24N/11$ & $24N/11$\tabularnewline
\hline 
$I_{3}$ &  & 11 & 11 & $9N/4$ & $9N/4$\tabularnewline
\hline 
$R_{2}I_{1}$ & r & 0.8 & 1.5 & $13N/4$ & $26N/4$\tabularnewline
\hline 
-''- & n & 0.4 & 0.75 & $13N/4$ & $13N/4$\tabularnewline
\hline 
$R_{1}I_{2}$ & r & 6 & - & - & $\infty$\tabularnewline
\hline 
-''- & n & 12 & - & $3N$ & $3N$\tabularnewline
\hline 
$I_{2}$ &  & 1 & 1 & $3N$ & $3N$\tabularnewline
\hline 
\end{tabular}

\caption{\label{tab:Combined-chains-}Combined chains $N_{comb}$ and order
parameters $m$ for different star connectivities for arms in ring
structures (r) or the remaining normal connections (n). \%$A$ and
\%$AB$ are the fractions of polymer that have the corresponding combined
chains. Fractions are simulation data at $c^{*}$, while combined
chains and order parameters are estimated using the phantom model
single defects in an otherwise ideal network structure.}
\end{table}

The results show that in particular for $AB$ networks there is a
single clearly distinct (a factor of 2 different) order parameter
for the most abundant non-ideal network structure close to $c^{*}$,
while the other most abundant defects ($I_{3}$ and ideal connections
of $R_{2}I_{2}$) have nearly non-distinguishable order parameter
as compared to the ideal connections. Note that all other structures
missing at the above table contribute each on order 1\% or less to
the fraction of polymer and thus, lead to a slight smearing out of
the full order parameter distribution. Thus, the structure $R_{2}I_{2}$
in $AB$ star networks is by far the most promising candidate for
investigating cyclic defects in polymer networks using NMR. 

\begin{figure}
\includegraphics[angle=270,width=0.8\columnwidth]{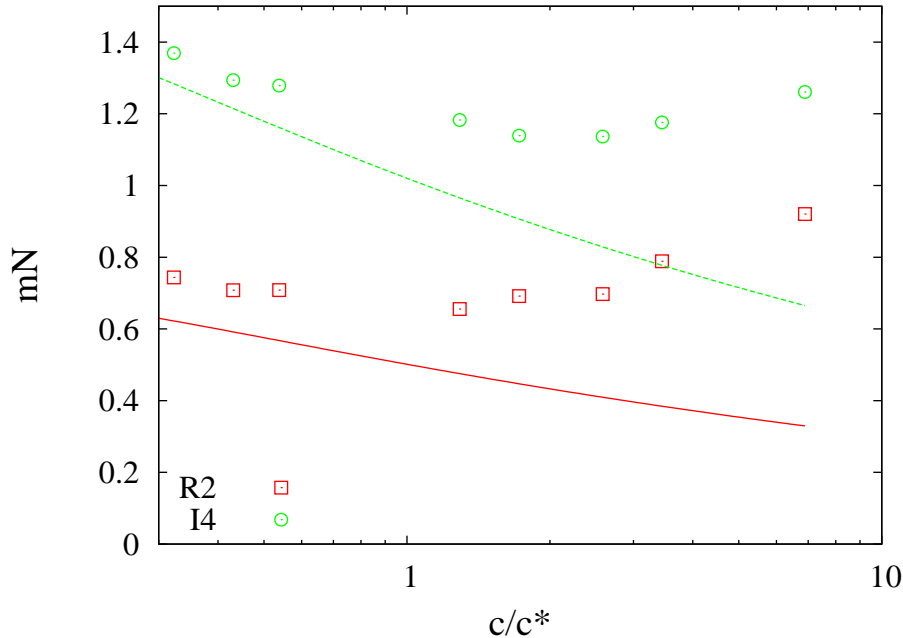}

\caption{\label{fig:Order-parameters-of}Scaled order parameters of $AB$ networks
with $N_{a}=32$ as function of concentration $c/c^{*}$ at cross-linking
compared with phantom model computations (lines) that include the
effect of chain swelling (see Appendix).}
\end{figure}

In the Appendix we additionally computed the effect of concentrations
onto the vector order parameters. In Fig. \ref{fig:Order-parameters-of}
we compare the order parameter as directly obtained in our simulations
with their values predicted for stars $I_{4}$ and in rings $R_{2}$
of stars $R_{2}I_{2}$. The data is multiplied by $N$ in order to
remove the chain length dependence of the phantom model. The limiting
behaviour without short loops is obtained by formally taking the limit
$c\rightarrow\infty$ (cf. table \ref{tab:Combined-chains-}). At
and below $c^{*}$ we find reasonable agreement between predictions
and simulation data. Note that both phantom model computations for
subsequent cross-linking and the extra excluded volume along the strands
after cross-linking extend the average chain conformation by roughly
10\%. For increasing concentrations, $c>c^{*}$, we find increasing
disagreement between data and theory. This difference can be explained
by the effect of entanglements, which lead to a scaling of $m\sim(NN_{e})^{-1/2}$
as shown in Ref. \cite{LangSommer10}. Thus, for the largest concentrations
there should be an increase of the order parameter as compared to
the phantom prediction by somewhat larger than a factor of two in
agreement with the simulation data. However, a sound analysis of the
concentration dependence of entanglements would require much larger
chain lengths and overlap numbers, since most data of the present
study is at the onset of entanglement effects. Note that the partial
compensation of entanglements and swelling effects leads to almost
unmodified order parameters for stars $I_{4}$ at $c/c^{*}$ slightly
larger than one. The effect of entanglements is much more pronounced
(due to the smaller order parameters at $c^{*})$ for the loops at
$R_{2}I_{2}$, which leads to an apparent exponent on the order of
order $0.2$ in the transition regime. This observation is in agreement
with recent experimental data \cite{lange2009diplomarbeit,Lange}
and will be elaborated with more detail in a forthcomming publication
using networks of clearly larger chain length.

\section{Conclusions}

We analyzed the structure of polymer networks obtained from star polymer
solutions for concentrations ranging from dilute to melt. AB-type
networks of symmetric composition, where crosslinking can occur only
between unlike species are compared with A-type networks, where crosslinking
takes place between all molecules (including self-links within a given
star). The analysis of the pair correlation function showed no essential
increase of the spatial order of stars in both types of networks upon
cross-linking beyond a weak nearest neighbor correlation. On intermediate
distances of the order of a few number of stars, concentration fluctuations
among $A$ and $AB$ networks are suppressed. Long range order could
not be detected. 

Network connectivity was analyzed in terms of the formation of short
ring structures (defects) that diminish the elastic response of the
network and might be detected in NMR-experiments. For $AB$-type networks,
double links between two neighboring stars, see Fig.~\ref{fig:Loop-structures-that-1},
are most abundant and their fraction is about 11\% at $c^{*}$. This
shows that $AB$-type networks are far from perfect in terms of connectivity.
The coincidence of the double links $R_{2}$ data of $AB$ networks
and the dangling loops $R_{1}$ and double links $R_{2}$ of $A$
networks after correcting the differences among both types of samples
indicates that effects of spatial order are ignorable for $AB$ networks.
The absence of $R_{1}$ structures leads to the formation of $AB$
networks with a significantly increased amount of active material
as compared to $A$ networks at same conditions. We argue that this
is the major difference between the two types of networks and is responsible
for an substantial increase in active material and, hence, for the
increased mechanical strength. 

Our study reveals that in particular for $AB$ networks there is only
one dominating (at concentrations close to $c^{*}$) short loop structure,
$R_{2}$, which has a clearly distinct segmental order parameter as
compared to most of the remaining network structure. Therefore, this
particular type of loops might lead to a distiguished signal in solid-state
NMR experiments as has been recently observed~\cite{Lange}. We observe
an apparent concentration dependence of the order parameters that
is clearly affected by entanglement effects at $c>c^{*}$. At or slightly
below $c^{*}$, the phantom model achieves a reasonable prediction
for the vector order parameter.

\section{Acknowledgement}

The authors thank the ZIH Dresden for a generous grant of computing
time. ML thanks the DFG for funding project LA2735/2-1. The authors
thank Ron Dockhorn and Marco Werner for stimulating discussions and
for assistance with the simulation tools.

\section{Appendix}

Let us assume that the network has ideal connectivity (no finite loops,
all junctions of functionality $f$ as used in section 7.2.2 of \cite{Rubinstein}
for deriving phantom modulus). If one strand is being removed of this
perfect network, the fluctuations of the cross-links previously attached
to this strand can be modeled by virtual chains of 
\begin{equation}
K=\frac{N}{f-2}\label{eq:K-1}
\end{equation}
monomers that are attached to the non-fluctuating elastic background.
When re-inserting the chain in between, we arrive at a combined chain
of 
\begin{equation}
N_{comb}=K+N+K=\frac{f}{f-2}N\label{eq:Ncomb-1}
\end{equation}
monomers that is fixed at both ends. Note that $\left[m_{k}\right]$
of this chain equals $N_{comb}^{-1}$.

Similarily one can show that removing two links leads to cross-link
fluctations as given by 
\begin{equation}
K=\frac{(f-1)}{(f-2)^{2}}N.\label{eq:K2-1}
\end{equation}
Inserting a double link in between two such cross-links leads to a
combined chain of
\begin{equation}
N_{comb}=2\frac{(f-1)}{(f-2)^{2}}N+N/2\label{eq:Ncomb2-1}
\end{equation}
monomers, since the double link is equivalent to a chain of $N/2$
monomers. However, each strand of the double link still contains $N$
monomers and the corresponding order parameter is reduced by an additional
factor of 2, because the average vector between the ends of this strand
is divided into twice as many sections. Thus, for $f=4$ we obtain
$N_{comb}=2N$ as for an ideal bond, but $\left[m\right]=1/(4N)$
instead of $\left[m\right]=1/(2N)$ for an ideal connection. Note
that the double link leads to increased fluctuations of the cross-links
attached (which reflect the local reduction of modulus) and thus,
affects the combined chains of the surrounding strands. To show this
effect, the combined chains and corresponding order parameters of
the directly connected surrounding chains were also computed and given
at table \ref{tab:Combined-chains-} (the chains at position n).

For applying the above computations to our simulation data, we have
to include the effect of concentration and distributions of functionalities
as function of concentration. Cross-linking at different concentration
affects first the equilibrium size of a network strand. As discussed
above, 
\begin{equation}
R_{e}^{2}\approx b^{2}\left(\frac{c}{c^{*}}\right)^{\frac{1-2\nu}{3\nu-1}}N^{2\nu}.\label{eq:2-1-1}
\end{equation}
One simple way to express the modified fluctuations of swollen chains
is to consider that these fluctuations are always equivalent in amplitue
to the size of the chains, since we discuss only samples at cross-linking
conditions. Thus, the virtual chains show the same {}``concentration
dependence'' as the real chains. Therefore,
\begin{equation}
m_{k}(c)\approx\frac{R^{2}}{b^{2}N^{2}}\approx\left(\frac{c}{c^{*}}\right)^{\frac{1-2\nu}{3\nu-1}}N^{2\nu-2}\label{eq:mk-1}
\end{equation}
for the combined chains of the phantom model. In consequence, conformational
changes upon cross-linking at different concentrations do not affect
the ratios (cf. table \ref{tab:Combined-chains-}) between the order
parameters of different structures, if the surrounding network structure
remains comparable.

Cross-linking at different concentrations also leads to a modification
of the weight average number of independent active connections. We
require the connectivity distribution to compute the weight average,
since the functionality of the connected neighbour is selected proportional
to its number of connections. Furthermore, only active connections
must be taken into account, since non-active parts of the network
do not contribute to the vector order parameter at $t\rightarrow\infty$.
Using the weight average functionality we implicitly assume that there
are no correlations between neighbouring functionalities, which is
clearly not the case (a double link always connects stars with reduced
functionality). But the results of a second study reveal \cite{Schwenke2},
that the effect of these correlations is ignorable in the vicinity
of $c^{*}$.

For the $AB$ networks of our study, the fraction of active junctions
(as given by star centers) and the fraction of active connections
among all existing connections is $>0.97$ for $c>c^{*}$. Thus, for
$c\gtrsim c^{*}$ we neglect a distinction between active and non-active
material and consider all existing connections as active. Note that
both quantities rapidly drop at concentrations clearly below $c^{*}$
and that the above approximation implicitly removes the small changes
in the average length of active strands by fixing it to $N$. Next,
we only distinguish between junctions (star centers) of three and
four connections, as justified by the data for $AB$ networks at $c^{*}$
in table \ref{tab:Combined-chains-}, and approximate that any double
link reduces the functionality of two junctions from four to three
at $c>c^{*}.$ Using this approximation and the best fit for loop
formation we find for the weight fraction of three functional junctions
approximately
\begin{equation}
w_{3}\approx4(1-p)+0.145\left(\frac{c}{c^{*}}\right)^{-0.53}\label{eq:w3-1}
\end{equation}
and consequently with 
\begin{equation}
w_{4}=1-w_{3}\label{eq:w4-1}
\end{equation}
the weight average functionality
\begin{equation}
f_{w}=\sum_{i=3}^{4}i^{2}w_{i}/\sum_{i=3}^{4}iw_{i}.\label{eq:fw-1}
\end{equation}
This weight average functionality replaces $f$ at equations (\ref{eq:K-1})
and (\ref{eq:K2-1}) and leads to increased average cross-link fluctuations
inside the sample for smaller concentrations. Thus, the contribution
of the virtual chains $K$ to the combined chains $N_{comb}$ has
an additional concentration dependence different to equation (\ref{eq:mk-1}).
This additional concentration dependence leads to a shift of the ratios
among the different order parameters as function of concentrations,
if the fraction of $K/N_{comb}$ is different for the particular structures.

Summarizing the above computations and approximations we find for
strands of $N$ monomers between two stars of type $I_{4}$ that 
\begin{equation}
N_{comb}\approx\frac{2(f_{w}-1)N}{(f_{w}-2)(f-1)}+N\label{eq:Nc-1}
\end{equation}
while for the double links inside $R_{2}I_{2}$ we obtain
\begin{equation}
N_{comb}\approx\frac{2(f_{w}-1)N}{(f_{w}-2)(f-2)}+N/2.\label{eq:Nc2-1}
\end{equation}
These results are inserted in equation (\ref{eq:mk-1}) to compute
the prediction for the order parameter as function of concentration
in Figure \ref{fig:Order-parameters-of}.

\bibliographystyle{plain}

\section*{\newpage{}}

\section*{Table of Contents Graphic}

\emph{Number fraction of cyclic defects}

Konrad Schwenke

Michael Lang

Jens-Uwe Sommer

\begin{figure}[htbp]
\includegraphics[angle=0,width=1\columnwidth]{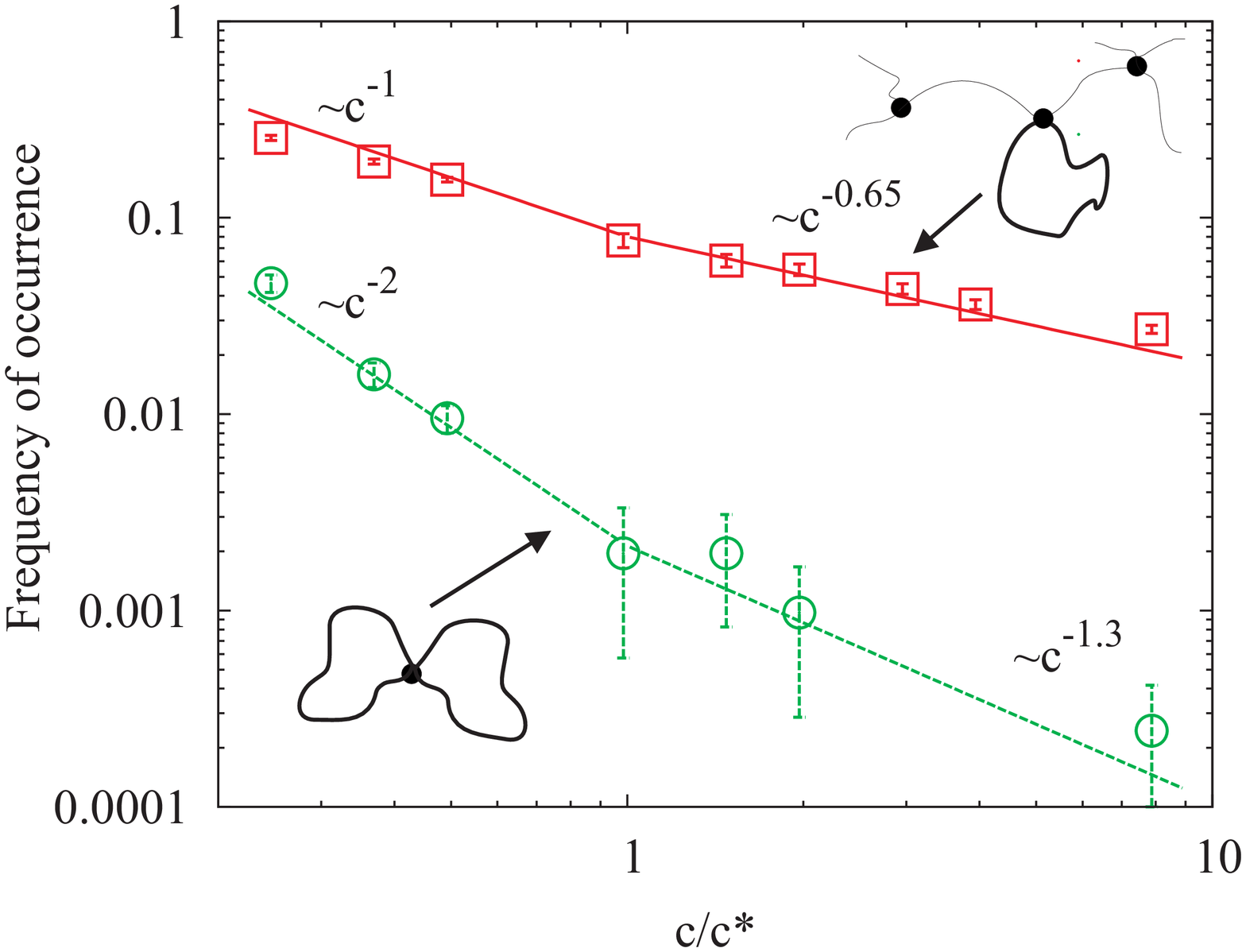}

\end{figure}

\end{document}